\documentclass[a4paper,11pt]{article}
\pdfoutput=1 
\usepackage{jheppub} 
\usepackage[english]{babel}
\usepackage{microtype,xspace}
\usepackage[utf8]{inputenc}	

\usepackage[capitalise]{cleveref}

\usepackage{amsfonts,amsmath,amssymb,bm,mathrsfs,mathtools,dsfont} 	
\allowdisplaybreaks 	
\usepackage{xcolor}
\usepackage{hyperref}
\usepackage{slashed}
\usepackage{pdflscape}

\usepackage{siunitx}
\usepackage{graphicx}
\graphicspath{{./Figures/}}
\usepackage{multirow}

\usepackage{tabularx}
\newcolumntype{Y}{>{\centering\arraybackslash}X}

\usepackage[shortlabels]{enumitem}
\setlist{itemsep=.1em,topsep=.5em}
\SetEnumerateShortLabel{i}{\textit{\roman*}}

\definecolor{red}{rgb}{0.6,.0706,.1373}
\definecolor{blue}{rgb}{0,0.396,0.741}
\colorlet{blueRef}{blue!80!black}
\colorlet{blueLink}{blue!80!black}
\hypersetup{
	colorlinks, 
	bookmarksopen, 
	bookmarksnumbered,
	citecolor=blueRef, 		
	linkcolor=blueLink,	
	urlcolor=blueLink,			
}

\DeclareMathVersion{sans}
\SetSymbolFont{operators}{sans}{OT1}{cmbr}{m}{n}
\SetSymbolFont{letters}{sans}{OML}{cmbrm}{m}{it}
\SetSymbolFont{symbols}{sans}{OMS}{cmbrs}{m}{n}
\SetMathAlphabet{\mathit}{sans}{OT1}{cmbr}{m}{sl}
\SetMathAlphabet{\mathbf}{sans}{OT1}{cmbr}{bx}{n}
\SetMathAlphabet{\mathtt}{sans}{OT1}{cmtl}{m}{n}
\SetSymbolFont{largesymbols}{sans}{OMX}{iwona}{m}{n}

\DeclareMathVersion{boldsans}
\SetSymbolFont{operators}{boldsans}{OT1}{cmbr}{b}{n}
\SetSymbolFont{letters}{boldsans}{OML}{cmbrm}{b}{it}
\SetSymbolFont{symbols}{boldsans}{OMS}{cmbrs}{b}{n}
\SetMathAlphabet{\mathit}{boldsans}{OT1}{cmbr}{b}{sl}
\SetMathAlphabet{\mathbf}{boldsans}{OT1}{cmbr}{bx}{n}
\SetMathAlphabet{\mathtt}{boldsans}{OT1}{cmtl}{b}{n}
\SetSymbolFont{largesymbols}{boldsans}{OMX}{iwona}{bx}{n}

\usepackage[noindentafter]{titlesec} 
\titleformat{\section}{\large \bfseries \sffamily \mathversion{boldsans} \color{blue!70!black} }{\thesection}{10pt}{}{}
\titlespacing{\section}{0pt}{10pt}{5pt}
\titleformat{\subsection}{\sffamily \mathversion{sans} \color{blue!80!black} }{\thesubsection}{10pt}{}{}
\titlespacing{\subsection}{0pt}{8pt}{3pt}
\titleformat{\subsubsection}{\normalsize \itshape \sffamily \mathversion{sans} \color{blue!80!black} }{\thesubsubsection}{10pt}{}{}
\titlespacing{\subsubsection}{0pt}{8pt}{2pt}

\renewcommand{\thesection}{\arabic{section}}
\renewcommand{\thesubsection}{\thesection.\arabic{subsection}}
\renewcommand{\thesubsubsection}{\thesubsection.\arabic{subsubsection}}
\makeatletter
\renewcommand{\p@subsection}{}
\renewcommand{\p@subsubsection}{}
\makeatother

\makeatletter
\let\MyIntOrig\int
\def\MyIntSpace{\hspace{-.35em}} 
\def\int{\MyInt}
\def\MyInt{\MyIntOrig\MyIntSkipMaybe}
\def\MyIntSkipMaybe{
	\@ifnextchar_{\MyIntSkipScript}{%
		\@ifnextchar^{\MyIntSkipScript}{%
			\@ifnextchar\limits{\MyIntSkipTok}{%
				\@ifnextchar\nolimits{\MyIntSkipTok}{%
					\MyIntSpace}}}}%
}
\def\MyIntSkipScript#1#2{#1{#2}\MyIntSkipMaybe}
\def\MyIntSkipTok#1{#1\MyIntSkipMaybe}

\newcommand{\pushright}[1]{\ifmeasuring@#1\else\omit\hfill$\displaystyle#1$\fi\ignorespaces}
\makeatother

\newcommand{\TeV}{\text{TeV}}
\newcommand{\GeV}{\text{GeV}}

\newcommand{\beq}{\begin{equation} }
\newcommand{\eeq}{\end{equation}}
\newcommand{\bi}{\begin{itemize} }
\newcommand{\ei}{\end{itemize} }

\allowdisplaybreaks

\begin{document}

\author[1]{Reuven Balkin,}
\author[2]{Eric Madge,}
\author[3]{Tony Menzo,}
\author[2]{Gilad Perez,}
\author[1]{Yotam Soreq,}
\author[3]{and Jure Zupan}

\affiliation[1]{Physics Department, Technion -- Israel Institute of Technology, Haifa 3200003, Israel}
\affiliation[2]{Department of Particle Physics and Astrophysics, Weizmann Institute of Science, Rehovot, Israel 7610001}
\affiliation[3]{Department of Physics, University of Cincinnati, Cincinnati, Ohio 45221,USA}

\emailAdd{reuven.b@campus.technion.ac.il}
\emailAdd{eric.madge-pimentel@weizmann.ac.il}
\emailAdd{menzoad@mail.uc.edu}
\emailAdd{gilad.perez@weizmann.ac.il}
\emailAdd{soreqy@physics.technion.ac.il}
\emailAdd{zupanje@ucmail.uc.edu}

\title{\mathversion{boldsans}
On the implications of positive $W$ mass shift
}

\abstract{
We investigate the phenomenological implications of the recent $W$ mass measurement by the CDF collaboration, which exhibits tension with the standard model~(SM) electroweak fit. 
Performing the fit to the electroweak observables within the SM effective field theory, we find that the new physics that contributes either to the determination of the electroweak vacuum expectation value, or to the oblique parameters, can improve the agreement with data. 
The best description is obtained from a fit where flavor universality is not required in the new physics operators, with  2 to 3 $\sigma$ indications for several nonzero Wilson coefficients. 
We point out that top partners with order TeV masses could lead to the observed shift in the $W$ mass.
}

\maketitle

\newpage

\section{Introduction}
\label{sec:intro}

Recently, the CDF collaboration provided a legacy measurement of the $W$ boson mass \cite{CDF:mW}
\beq
    \label{eq:mWCDF}
    m^{\rm CDF}_W=\SI{80.4335\pm0.0094}{\GeV}\,.
\eeq
This most precise measurement of $m_W$ represents a shift of $3.6\,\sigma$ compared to the 2021 PDG average of  the LEP~\cite{ALEPH:2013dgf}, ATLAS~\cite{ATLAS:2017rzl} and the previous Tevatron combination~\cite{CDF:2013dpa}, which yields $m^{\rm PDG}_W=\SI{80.379\pm0.012}{\GeV}$~\cite{ParticleDataGroup:2020ssz}.
Similarly, a naive average of LEP, ATLAS and the LHCb~\cite{LHCb:2021bjt} measurement leads to $m^{\rm NWA}_W=\SI{80.368\pm0.014}{\GeV}$, a difference of $3.8\,\sigma$ from \cref{eq:mWCDF}, showing that there is significant tension between the CDF measurement and the determinations of $m_W$ by other experiments. 
Working under the assumption that the CDF measurement will be confirmed by future measurements, we explore its phenomenological implications.  
Intriguingly, the CDF measurement is also in $\sim 7.0\,\sigma$ tension with the Standard Model~(SM) prediction from a global electroweak fit, $m_W ^{\rm fit}=\SI{80.361\pm0.006}{GeV}$~\cite{ParticleDataGroup:2020ssz}. 
In this paper, we perform several fits to $Z$ and $W$ pole observables with varying assumptions on new physics: just the operators that modify the determination of the electroweak vacuum expectation value~(VEV), new physics~(NP) only in the oblique corrections, or the general fit that allows also for non-universal couplings of new physics. 
We then interpret the results of the fits in terms of several simple new physics explanations that could bring the $W$ measurement by CDF in agreement with the predictions from the electroweak fit. 
For other recent discussions of the phenomenological implications of the CDF $m_W$ measurements see Refs.~\cite{Lu:2022bgw,Strumia:2022qkt,deBlas:2022hdk,Fan:2022yly,Tang:2022pxh,Cacciapaglia:2022xih,Blennow:2022yfm,Liu:2022jdq,Lee:2022nqz,Cheng:2022jyi,Song:2022xts,Bagnaschi:2022whn,Paul:2022dds,Bahl:2022xzi,Asadi:2022xiy,DiLuzio:2022xns,Athron:2022isz,Gu:2022htv,Babu:2022pdn,Arias-Aragon:2022ats,Sakurai:2022hwh,Athron:2022qpo,Heckman:2022the}.\footnote{The quantitative analysis of whether or not these follow the universal scaling laws uncovered in \cite{Backovic:2016xno} we defer to future work, at which point we anticipate more data to become available.}

The paper is organized as follows. 
In \cref{sec:EFT} we perform a general analysis of the electroweak data, including the new CDF measurement of $m_W$. 
\Cref{sec:models} contains a discussion of sample new physics models that can alleviate the tension of the SM predictions with data, while \cref{sec:conc} contains our conclusions. 
Further details about the fit are collected in \cref{sec:oldFit,app:inputs}.

\section{General analysis}
\label{sec:EFT}

In performing the global fit to electroweak observables, allowing for dimension six new physics operators, we follow the approach of Refs. \cite{Falkowski:2014tna,Efrati:2015eaa} (see also~\cite{Falkowski:2019hvp,Breso-Pla:2021qoe} as well as the early work in \cite{Grinstein:1991cd}). 
First the SM electroweak fit is performed, the results of which are then used to obtain the constraints on the Wilson coefficients of dimension six operators 
\beq
    {\cal L}_{\rm eff}
=   {\cal L}_{\rm SM}+\frac{1}{v^2}\sum_i c_i {\cal O}_{6,i},
\eeq
where $v$ is the Higgs VEV and the sum in the second term runs over the SM effective field theory~(SMEFT) operators in the Warsaw basis~\cite{Grzadkowski:2010es}. 
The global $\chi^2$ function that is used to obtain the bounds on the SMEFT Wilson coefficients is given by, 
\beq
    \chi^2
=   \sum_{i,j}\big(
    O_{i,{\rm exp}}-O_{i,{\rm th}}\big)\sigma_{ij}^{-2}
    \big(O_{j,{\rm exp}}-O_{j,{\rm th}}\big) \, ,
\eeq
with $\sigma_{ij}^{-2}$ the experimental covariance matrix, $O_{i,{\rm exp}}$ the experimental meausurement of the particular electroweak observable, while its theory prediction at next-to-next-to-leading order in the SM and expanded to linear order in new physics is
\beq
    O_{i,{\rm th}}
=   O_{i,{\rm SM}}^{\rm NNLO}+\vec \delta g 
    \cdot O_{i,{\rm NP}}^{\rm LO} \, .
\eeq
For the SM we use the latest theory results, with the numerical values collected in Ref.~\cite{Breso-Pla:2021qoe} based on~\cite{Falkowski:2019hvp,dEnterria:2020cpv,Awramik:2003rn}, while for NP corrections it suffices to calculate to leading order. 
The 19 pole parameters $\delta g_i$ depend on the dimension 6 Wilson coefficients, with the dependence given in Appendix A of Ref.~\cite{Efrati:2015eaa}, while the observables used in our fit are listed in \cref{tab:obs}.    
The NP corrections to the observables depend on the NP Wilson coefficients as well as on the two weak gauge couplings and the electroweak vacuum expectation values,
\begin{equation}
    g_L = 0.6458\,,\qquad 
    g_Y = 0.3580\,,\qquad 
    v = \SI{246.22}{\GeV} \,,
\end{equation}
which are obtained by using the LO SM relations for the Fermi constant $G_F=1/(\sqrt{2} v^2) $, the $Z$ boson mass,  $m_Z = \sqrt{g_L^2 + g_Y^2} v/2 $, and the fine structure constant at $m_Z$, $\alpha(m_Z)^{-1} = 4 \pi (g_L^2 + g_Y^2)/(g_L^2 g_Y^2)$. 
The SM relations get corrected by the NP contributions, which is taken into account when writing the expressions for  $O_{i,{\rm NP}}^{\rm LO}$ in terms of the dimension 6 Wilson coefficients. 
Setting all NP contributions to zero gives for the SM $\chi^2_{\rm SM} = 99.3$ with 42 observables contributing to the $\chi^2$.

In light of the new CDF measurement of $m_W$, it is interesting to focus on the NP that can directly result in a shift in the $W$ mass relative to the prediction for the $Z$ mass. 
Only five dimension 6 operators result in such a shift
\beq
    \label{eq:mW}
    \delta m_W
    \equiv
    \frac{m_W\big|_{\rm exp}-m_W\big|_{\rm SM}}{m_W}
    =\frac{1}{g_L^2-g_Y^2}\Big( -g_L^2 g_Y^2 c_{WB}+g_L^2 c_T-g_Y^2 \delta v\Big),
\eeq 
where the normalized shift in the electroweak VEV due to NP is given by\footnote{We note the typographical error in \cite{Efrati:2015eaa} in the sign multiplying $[c_{\ell \ell}]_{1221}$.}
\beq
    \label{eq:deltav}
    \delta v
=   \frac{1}{2} \big([c_{H\ell}']_{11}+[c_{H\ell}']_{22}\big)
    -\frac{1}{4}[c_{\ell \ell}]_{1221}.
\eeq
The corresponding operators are given by
\begin{equation}\begin{aligned}
    {\mathcal O}_T
    &=\big(H^\dagger  \stackrel{\leftrightarrow}{D_\mu} H)^2,  
    &{\mathcal O}_{\rm WB}
    &=g_L g_Y H^\dagger \sigma^i H W_{\mu\nu}^i B_{\mu\nu},
\\
    {\mathcal O}_{H\ell}'
    &=(\bar \ell \sigma^i \gamma_\mu \ell)
    \big(H^\dagger  i \sigma^i \stackrel{\leftrightarrow}{D_\mu} H),  
    & {\mathcal O}_{\ell\ell}
    &=(\bar \ell  \gamma_\mu \ell)(\bar \ell  \gamma^\mu \ell),
\end{aligned}\end{equation}
where $\sigma^i$ acts on the $SU(2)_L$ space and we suppressed the generational indices.

\subsection{New physics in the muon decay}
\label{sec:muon:dec:fit}

We first explore the possibility that the NP modifies the muon lifetime, from which the Fermi constant is extracted. 
The main observation here is that in the electroweak fits the SM predictions for the $W$ mass is obtained from its relation to the Fermi constant, $G_F$, which is very precisely determined from muon lifetime. 
A change to the muon lifetime due to new physics contributions then directly translates to a shift in the prediction for $m_W$ from the electroweak fit~\cite{Capdevila:2020rrl,Crivellin:2021njn}. 

In the SM electroweak fit the $W$ mass is predicted from the SM relation~\cite{Awramik:2003rn}
\beq
    \label{eq:mWfit}
    m_W^2|_{\rm fit}
=   \frac{m_Z^2}{2}\left[1+\left(1-
    \frac{\sqrt 8 \pi \alpha (1-\Delta r)}{G_F m_Z^2}\right)^{1/2}\right],
\eeq
where $m_Z$ and $G_F$ are the measured $Z$ boson mass and Fermi constant, $\alpha$ is the fine structure constant, and $\Delta r$ are the radiative corrections that depend on the other SM parameters, including the $m_W$. 
If the total decay width of the muon, $\Gamma_\mu$, receives a correction $\Delta \Gamma_\mu^{\rm NP}$ from NP, 
\beq
    \Gamma_\mu
=   \Gamma_\mu^{\rm SM}+\Delta \Gamma_\mu^{\rm NP},
\eeq
this shifts the value of $G_F$ away from the SM, and translates to a modification of the predicted $m_W$ 
\beq
    \delta m_W
=   -\frac{1}{4}\frac{m_Z^2-m_W^2}{2 m_W^2-m_Z^2}\frac{\Delta \Gamma_\mu^{\rm NP}}{\Gamma_\mu^{\rm SM}}.
\eeq
This is a special case of the more general expression, \cref{eq:mW},
 assuming here only a NP change to $\delta v$, 
\beq
    \frac{\Delta \Gamma_\mu^{\rm NP}}{\Gamma_\mu^{\rm SM}}
=   4 \delta v={2} \big([c_{H\ell}']_{11}+[c_{H\ell}']_{22}\big)
    -[c_{\ell \ell}]_{1221}.
\eeq
Note that explaining the CDF $m_W$ measurement requires a negative interference with the SM in the muon decay width,
\beq
\label{eq:DeltaGamma:mw}
    \frac{\Delta \Gamma_\mu^{\rm NP}}{\Gamma_\mu^{\rm SM}}
    =(-9.0\pm1.3)\times 10^{-13}\,.
\eeq
While in principle $\Gamma_\mu$ can be modified by new muon decay channels, potentially with light new physics in the final state, such contributions would always result in an increased $\Gamma_\mu$. 
A possible explanation of the CDF $m_W$ measurement thus requires an interference with the SM in the $\mu \to e \nu_\mu \bar \nu_e$ decay. 
If this comes from negative interference with a $W'$ this implies $m_{W'}\simeq \SI{1.2}{\TeV}$ for $g'=g$ but opposite sign of the couplings to either the muon or electron coupling, compared to the ones for the SM $W$. 
The numerical value in \cref{eq:DeltaGamma:mw} assumes that the NP that contributes to $\Gamma_\mu$ does not affect any other observables, which is not true for heavy new physics that matches onto SMEFT. 
We therefore perform a global fit taking all such correlations between different electroweak observables into account. 

\begin{figure}
    \centering
    \includegraphics[width=0.8\textwidth]{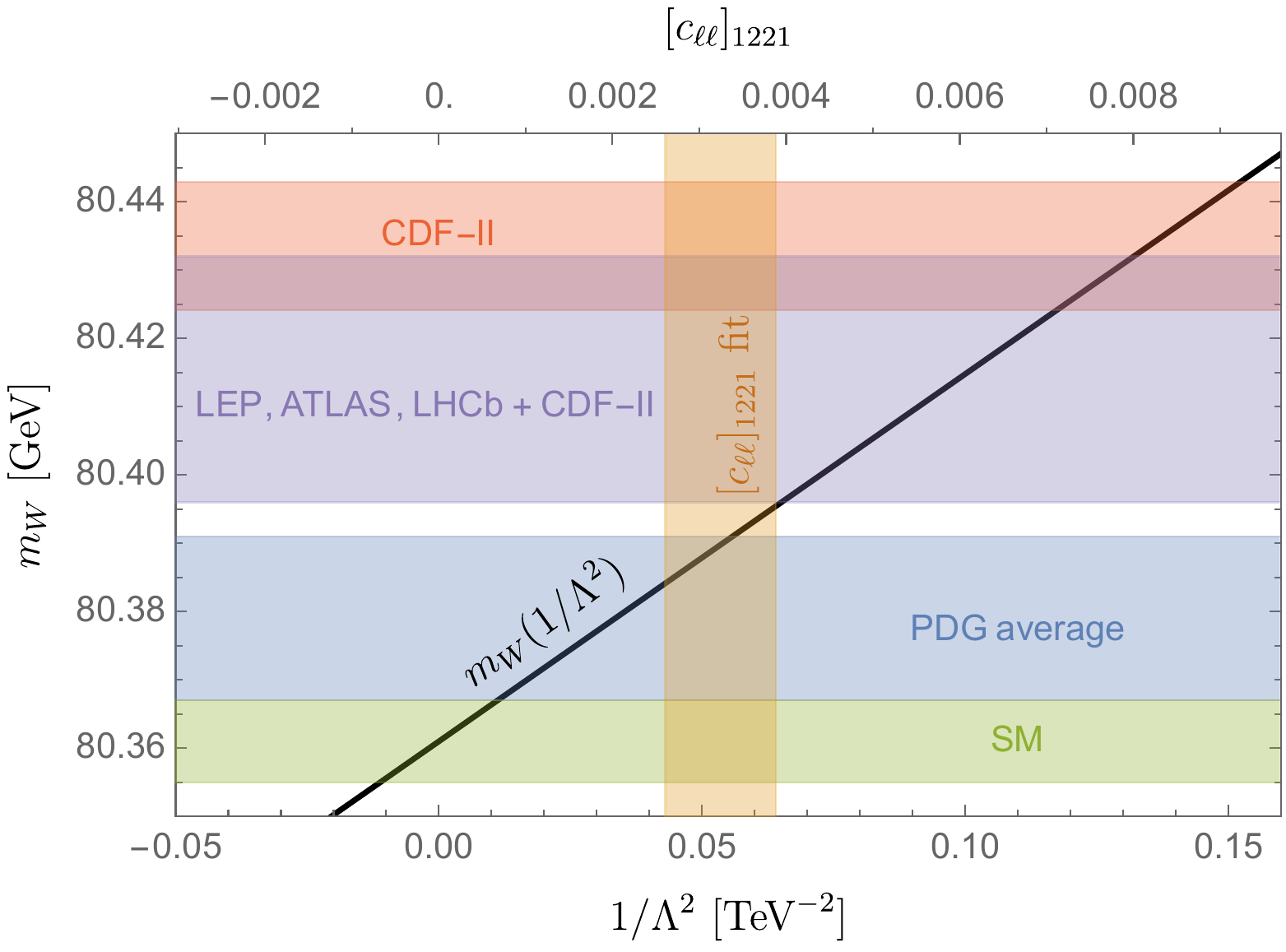}
    \caption{The results of an electroweak fit, where only $1/\Lambda^2 = [c_{\ell\ell}]_{1221}/v^2$ is allowed to be nonzero (orange band). The horizontal bands show the $1\sigma$ range of $m_W$ measurement: the PDG average (blue), the recent CDF measurement (red), and our naive average of the two (magenta), along with the prediction of the SM electroweak fit (green). The black line shows the dependence of $m_W$ on $1/\Lambda^2$.}
    \label{fig:mWcll}
\end{figure}

Assuming that the NP resides only in the  $[c_{\ell\ell}]_{1221}$ operator and setting all other Wilson coefficients to zero, the fit to the electroweak observables gives
\begin{equation}
    [c_{\ell\ell}]_{1221} = (3.2\pm0.6)\times 10^{-3}\,,\qquad 
\end{equation}
corresponding to
\begin{align}
    m_W = \SI{80.390(6)}{\GeV}\,.
\end{align}
and $\chi^2_{\rm SM} - {\rm min}(\chi^2)=25.9$, which is a significant improvement. 
However, the goodness of fit is still relatively poor $\chi^2/{\rm d.o.f.}=73.4/(42-1)$. 
This can be traced to the fact that the change in $\Gamma_\mu$ cannot explain fully the CDF $m_W$ measurement, since the change in the extracted electroweak vacuum expectation value, $\delta v$, also feeds in to many of the electroweak observables. 
This is illustrated in Fig.~\ref{fig:mWcll}, where the black solid line denotes the predicted shift in $m_W$ as a function of $1/\Lambda^2\equiv [c_{\ell \ell}]_{1221}/v^2$. 
The $1\sigma$ best fit band for $[c_{\ell \ell}]_{1221}$ is shown as an orange band. 
We see that a nonzero $[c_{\ell \ell}]_{1221}$ can increase $m_W$ somewhat above the SM prediction (green), even to the upper range of the PDG average of $m_W^{\rm PDG}$ measurements (blue) which does not yet contain the CDF measurement. 
It cannot, however, give a shift fully consistent with the CDF measurement (red band). 
For convenience, we also show the naive average of LEP, ATLAS, LHCb and the new CDF measurement for $m_W$ (magenta), where the errors were inflated according to the PDG prescription, giving   $m_W^{\rm inf}=80.414\pm0.018\,\GeV$.

\subsection{New physics modifying the extraction of the electroweak VEV}
\label{sec:vev:fit}

To capture the effects of other types of new physics that can modify the determination of the electroweak VEV, we perform a fit including all Wilson coefficients entering $\delta v$, cf.~\cref{eq:deltav}, while setting the remaining Wilson coefficients in the SMEFT Lagrangian to zero.
This yields
\begin{equation}
    \begin{pmatrix}
        [c_{\ell\ell}]_{1221} \\
        [c'_{H\ell}]_{11} \\
        [c'_{H\ell}]_{22}
    \end{pmatrix} = 
    \begin{pmatrix}
        -0.9\pm1.8 \\
        -1.1\pm0.4 \\
        -1.4\pm0.7
    \end{pmatrix} \times 10^{-3} \,,\qquad
    \rho = \begin{pmatrix}
        1 & 0.48 & 0.89 \\
        0.48 & 1 & 0.25 \\
        0.89 & 0.25 & 1
    \end{pmatrix}\,,
\end{equation}
with a minimum $\chi^2=63.8$ and a prediction for the $W$ mass of $m_W = \SI{80.396\pm0.006}{\GeV}$. That is, if all three parameters entering $\delta v$ are allowed to vary in the fit, the $ [c_{\ell\ell}]_{1221}$ is found to be consistent with zero, and the shift in $m_W$ is due to $[c'_{H\ell}]_{ii}$.

\subsection{A fit to the $S$ and $T$ parameters}

\begin{figure}
    \centering
    \includegraphics[width=0.8\textwidth]{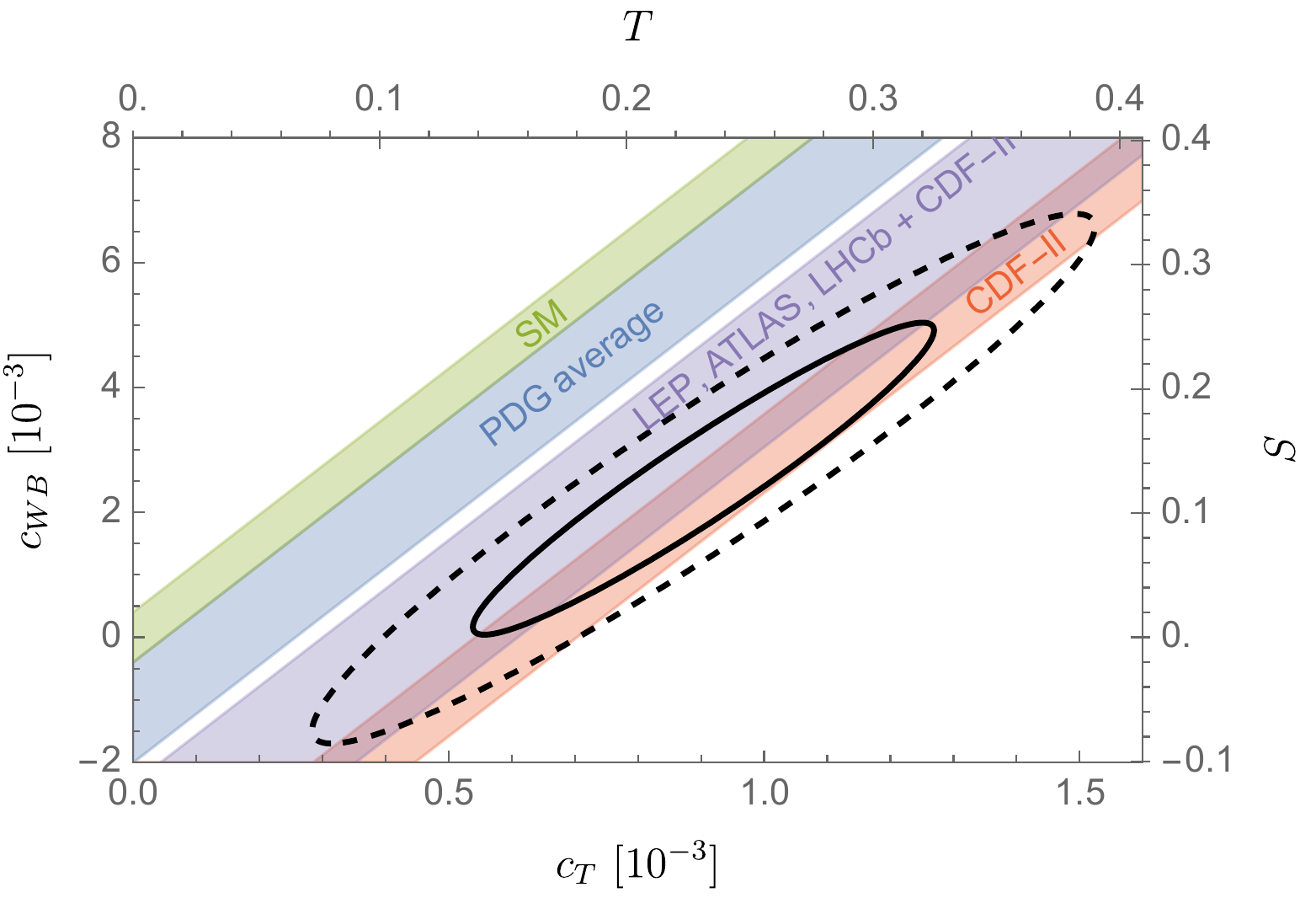}
    \caption{The fit results for coefficients $c_T$ and $c_{WB}$ along with expectation bands for different $m_W$ measurements: SM (green), PDG average (blue), naive average (purple), and the new results from CDF (red). The $1\sigma$ (solid) and $2\sigma$ (dashed) contours indicate the values from the fit. 
    }
    \label{fig:cWBcT}
\end{figure}

The oblique parameters $S$ and $T$ encode NP contributions via gauge-boson vacuum-polarization corrections~\cite{Peskin:1990zt}. 
We assume at the scale $m_Z$ the only non-negligible coefficients are given by $c_{WB}$ and $c_T$ in alignment with many NP models. 
In the Warsaw basis the coefficients $c_{WB}$ and $c_T$ are related to $S$ and $T$ (assuming $U=0$) via the relations, see, e.g.~\cite{Ellis:2018gqa,Falkowski:2019hvp}
\begin{equation}
         c_{WB} = \frac{1}{16\pi }S, 
         \hspace{0.2in} 
         c_T = \frac{g^2_L g^2_Y}{8\pi (g_Y^2 + g^2_L)}T.
\end{equation}
By fitting in the two-dimensional space we find 
\begin{equation}\label{ST}
        c_{WB} = (2.5\pm1.6)\times 10^{-3} \,, 
        \hspace{0.2in} 
        c_T = (9\pm2)\times 10^{-4} \,, 
\end{equation}
with a correlation coefficient $\rho = 0.95$ and the minimum $\chi^2=42.8$, and thus the goodness of fit $\chi^2/{\rm d.o.f.}=42.8/(42-2)$ in line with expectations for a valid description of data. 
The corresponding $W$ mass is $m_W = \SI{80.428\pm0.009}{\GeV}$. 
The above values of $c_{WB}, c_T$ translate to 
\beq
    S = \num{0.13(8)}, \qquad T = \num{0.23(5)},
\eeq
in agreement with current constraints on the $S$ parameter from $h \rightarrow \gamma \gamma$ \cite{Ellis:2018gqa}. 
In Fig.~\ref{fig:cWBcT} we show the results of our fit including estimated $S$ and $T$ values.

\subsection{A global flavorful fit}

Finally, we vary all the $Z$ and $W$ couplings to preform a global flavorful fit, following Refs.~\cite{Efrati:2015eaa,Falkowski:2019hvp}. 
In particular, this fit also allows for generational dependence of $W$ and $Z$ couplings to the SM fermions. 
We quote the results in  in term of the modified Wilson coefficients in the Warsaw basis, see Appendix~A of~\cite{Efrati:2015eaa},
\begin{gather}
    [c_{\ell\ell}]_{1221} = (-1.31\pm 0.90)\times 10^{-2}\,, \notag\\
    [\hat{c}'_{H\ell}]_{ii} = \begin{pmatrix}
         -0.62\pm 0.32 \\
         -0.45\pm 0.26 \\
         -0.06\pm 0.41
    \end{pmatrix} \times 10^{-2}\,,\qquad 
    [\hat{c}_{H\ell}]_{ii} = \begin{pmatrix}
         0.27\pm 0.31 \\
         0.04\pm 0.34 \\
         -0.31\pm 0.42 
    \end{pmatrix} \times 10^{-2}\,, \notag\\
    [\hat{c}_{He}]_{ii} = \begin{pmatrix}
         -0.09\pm 0.06 \\
         -0.17\pm 0.27 \\
         -0.30\pm 0.13 
    \end{pmatrix} \times 10^{-2}\,, \label{eq:fit:results:flavor}\\
    [\hat{c}'_{Hq}]_{ii} = \begin{pmatrix} 
         0.41\pm 2.67 \\
         -1.58\pm 2.78 \\
         -0.49\pm 4.10 
    \end{pmatrix} \times 10^{-2}\,,\qquad 
    [\hat{c}_{Hq}]_{ii} = \begin{pmatrix}
         3.08\pm 5.49 \\
         -0.91\pm 2.83 \\
         -0.42\pm 4.11 
    \end{pmatrix} \times 10^{-2}\,, \notag\\
    [\hat{c}_{Hu}]_{ii} = \begin{pmatrix}
         1.05\pm 6.27 \\
         0.84\pm 1.04 \\
        \text{---}
    \end{pmatrix} \times 10^{-2}\,,\qquad 
    [\hat{c}_{Hd}]_{ii} = \begin{pmatrix}
         5.07\pm 25.84 \\
         -6.46\pm 9.68 \\
         -4.54\pm 1.74 
    \end{pmatrix} \times 10^{-2}\,.\notag
\end{gather}
This corresponds to a $\chi^2 = 26.6$ and a $W$ mass of $m_W=\SI{80.434(9)}{\GeV}$ at the best-fit point. The goodness of fit $\chi^2/{\rm d.o.f.}=26.6/(42-21)$ is in fact better than one would naively expect. 

\begin{figure}
    \centering
    \includegraphics[width=.95\textwidth]{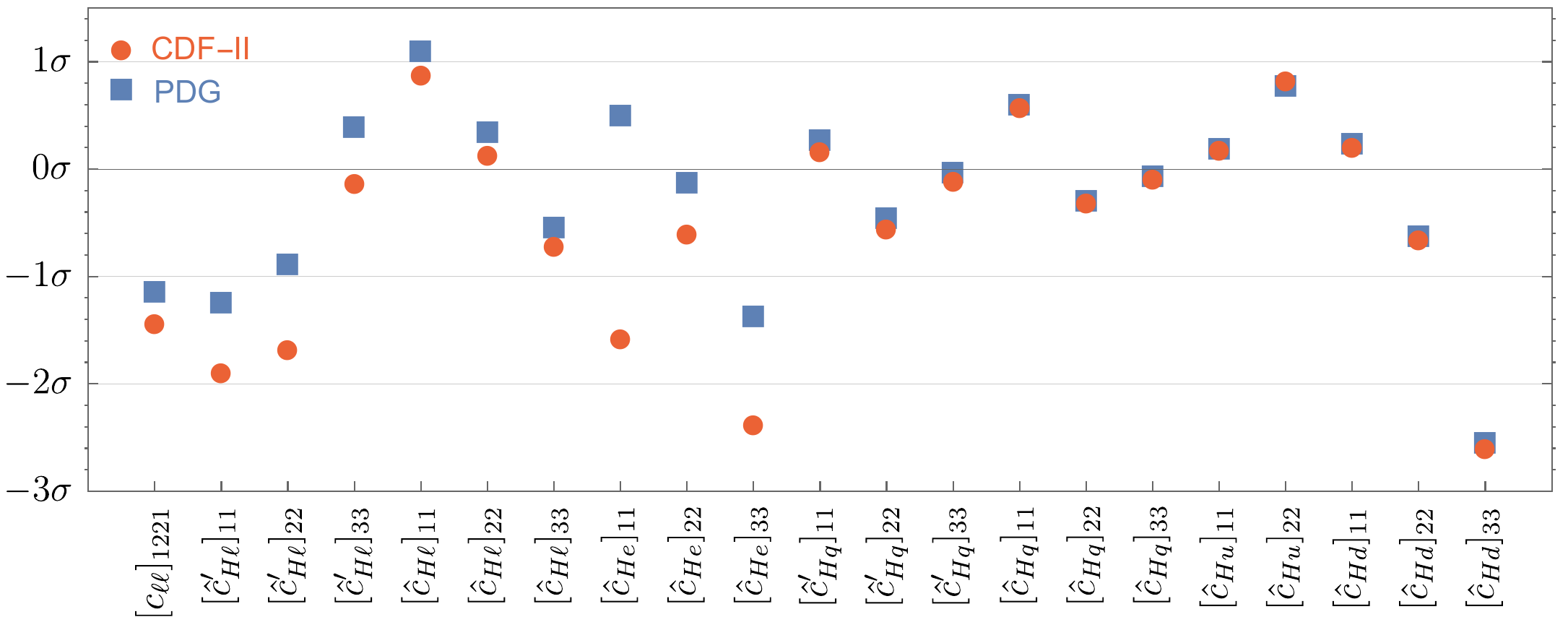}
    \caption{Pulls of the flavorful fit results using the CDF-II measurement of the $W$ mass (red dots) and PDG average (blue rectangles).}
    \label{fig:pull}
\end{figure}

The pulls for the values of the modified Wilson coefficients \eqref{eq:fit:results:flavor} are plotted in \cref{fig:pull} (red dots), where we compare them to the pulls that are obtained from a global fit in which the only change in the inputs is the use of the PDG average for the $W$ mass, $m_W^{\rm PDG}$, see also \cref{sec:oldFit}.
Note that there is no single operator that exhibits a $3\sigma$ evidence for a nonzero value, both within the electroweak fit with the PDG as well as in the fit with the CDF value of $m_W$. A comparison of the two fit results shows that there are a number of effective Wilson coefficients in which the evidence for nonzero value strengthened. Most notably, these are the effective Wilson coefficients that contribute to $\delta v$, the coefficients  $[c_{\ell\ell}]_{1221}$, $[\hat{c}^\prime_{H\ell}]_{11}$, $[\hat{c}^\prime_{H\ell}]_{22}$, where
\beq
  [\hat{c}^\prime_{H\ell}]_{ij}=  [{c}^\prime_{H\ell}]_{ij}+\delta_{ij}\biggr(g_L^2 c_{WB}-\frac{g_L^2}{g_Y^2} c_T\biggr).
\eeq
The flavor universal part is given by a linear combination of  $S$ and $T$ parameters, while the fit suggests also nonzero flavor non-universal contributions. A large change in the pull is also observed in the effective Wilson coefficient $[\hat{c}_{He}]_{11}$ and $[\hat{c}_{He}]_{33}$, where
\begin{align}
 [\hat{c}_{He}]_{ij}=    [{c}_{He}]_{ij} -2 c_T \delta_{ij},
\end{align}
Also in this case the data shows preference for nonvanishing flavor non-universal contributions.


\section{Models}
\label{sec:models}

\subsection{Tree level contributions to $\delta v$}
The two models that can lead to a tree level contribution to $[c_{\ell \ell}]_{1221}$ are the $W'$ mediator and the $S_1$ scalar mediator, which we discuss in turn.

\paragraph{The $W'$ with effective interactions.} A heavy vector boson, $W_\mu'^a$, a triplet of $SU(2)_L$, can have the following effective interactions with the leptons
\beq
{\cal L}_{W'}\supset \frac{1}{2} W_\mu'^a \Big( g_H H^\dagger \sigma^a i  \stackrel{\leftrightarrow}{D_\mu} H +g_{\ell_i} \bar \ell_i \sigma^a \gamma_\mu \ell_i+\cdots\Big).
\eeq
Integrating out the $W'$ gives
\begin{align}
[c_{\ell \ell}]_{1221}&=- \frac{g_{\ell_1} g_{\ell_2}}{2}\frac{v^2}{m_{W'}^2}=\num{3e-3} \times \big(-g_{\ell_1} g_{\ell_2}\big)\biggr(\frac{\SI{3.1}{\TeV}}{m_{W'}}\biggr)^2,\\
 [c_{H\ell}']_{ii}&=-\frac{g_H g_{\ell_i}}{4 m_{W'}^2}=-10^{-3} \times \big(g_{H} g_{\ell_i}\big)\biggr(\frac{\SI{3.9}{\TeV}}{m_{W'}}\biggr)^2,
\end{align}
where $m_{W'}$ is the $W'$ mass. 
In the numerical examples on the r.h.s.\ we used the values close to the central values of the fits in \cref{sec:muon:dec:fit,sec:vev:fit}. 
If the shift in $m_W$ is explained by just new physics in $[c_{\ell \ell}]_{1221}$ this needs to be positive, see \cref{sec:muon:dec:fit}, which would mean that the $W'$ needs to couple to the first and the second generation fermions with the opposite sign. 
Since the couplings of the new vector boson are then flavor non-universal, this is not easy to achieve in a complete UV model, while satisfying all the flavor constraints. 
In contrast, when all three operators are allowed to be varied in the fit, the main contribution is found to arise from $[c_{H\ell}']_{ii}$, see \cref{sec:vev:fit}. 
The pattern in the data can thus be matched if $g_H$ is larger than $g_{\ell_i}$. 

\paragraph{Tree level contribution from charged scalar. }
The scalar $S_1$ that carries a hypercharge~$1$ has couplings to lepton doublets of the form~\cite{deBlas:2017xtg}
\beq
    {\cal L}_{S_1}
    \supset  
    y_{ij} S_1\bar \ell_{Li} i \sigma_2 \ell_{Lj}^c +{\rm h.c.}.
\eeq
Integrating out the scalar generates the $[c_{\ell\ell}]_{1221}$ from the couplings to the first two generations
\beq
    [c_{\ell\ell}]_{1221}
    =-\frac{y_{12}y_{12}^*}{2}\frac{v^2}{m_{S_1}^2}.
\eeq

\subsection{Top-partners}

\begin{figure}
    \centering
    \includegraphics[width=.75\textwidth]{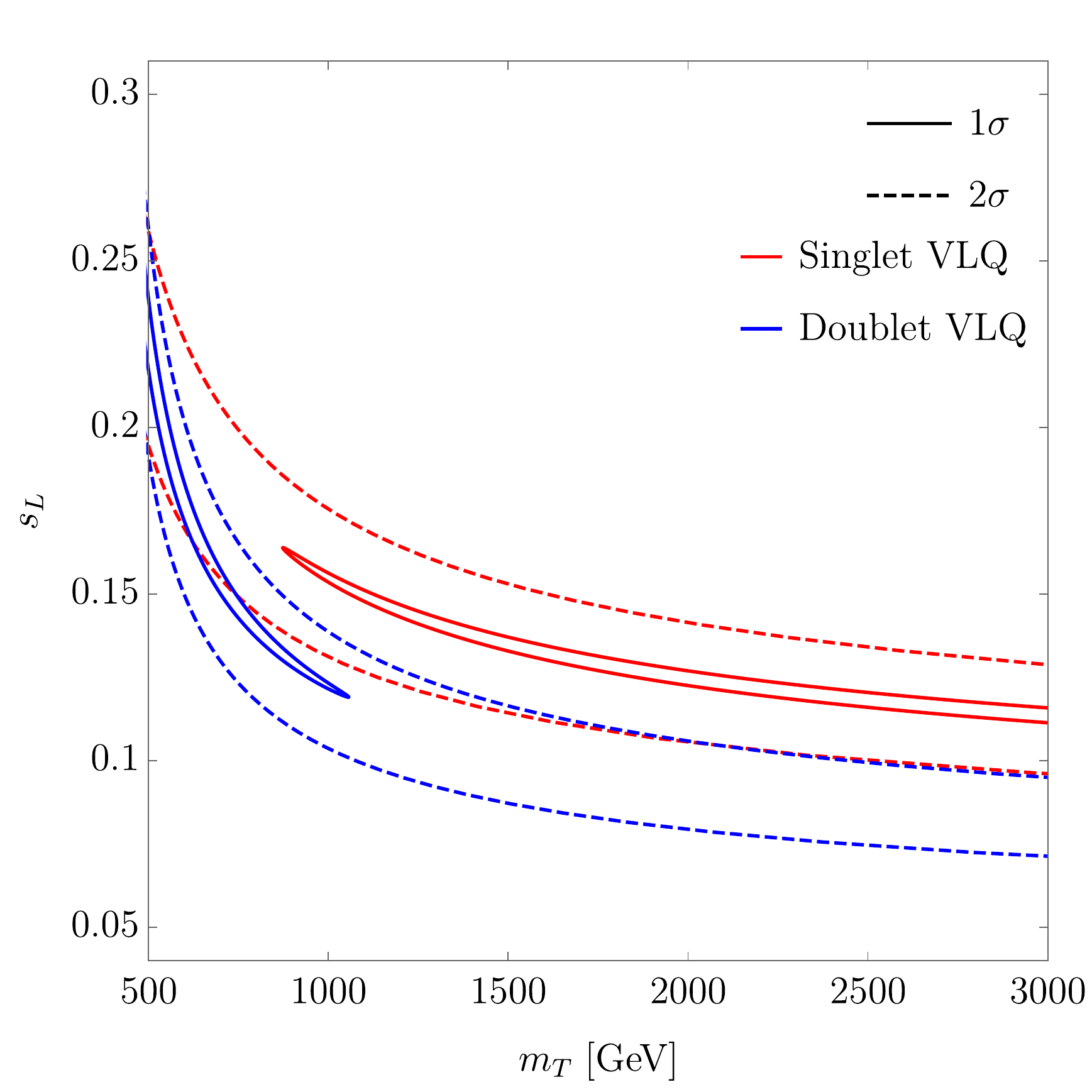}
    \caption{ The one (solid) and two (dashed) sigma CL contours for the singlet (red) and doublet (blue) partners where on the vertical axis we show the mixing-angle as a function of the vector-like (VLQ) quark mass, see text for more details.}
    \label{partners}
\end{figure}

Naturalness demands that the quadratic divergence of the one-loop top
contribution to the Higgs mass is cancelled at roughly the TeV scale. 
In many of the known extensions of the SM this is achieved, mechanically, by adding top partners, scalars or fermions, weak-singlets and doublets with order one Yukawa couplings to the SM top sector (see for instance Ref.~\cite{Berger:2012ec} for a more detailed discussion). 
 Here we focus on the question whether fermionic, spin-1/2, top partners can give rise to the relevant shifts in the oblique parameters found in \cref{ST}, such that the CDF measurement of the $W$ mass is explained, and set aside the precise details of the full UV model.
 For concreteness we consider the minimal content, adding either a single vector-like top-singlet partner or a vector-like third-generation doublet partner. 
 In Fig.~\ref{partners} we show the one and two sigma CL contours for the singlet and doublet partners where on the vertical axis we show the mixing-angle as a function of the partner mass.
 In partner models we expect that the mixing angle is of the order of $v/\sqrt2 m_T$, with $m_T$ the partner mass (for simplicity, we consider a single mass and a single mixing angle, in the heavy mass limit as required by the LHC limits~\cite{Banerjee:2022xmu}, for a complete description of the dependencies see, {\it e.g.}, Ref.~\cite{Chen:2017hak}).
 We find that models with such spin-1/2 partners can account for the observed shift in the $W$ mass, with an order one Yukawa, for a broad range of parameter space for TeV partner-masses.

\section{Conclusions}
\label{sec:conc}

While tension between the CDF legacy  measurement of the $W$ mass and the global average of $m_W$ determinations may give one pause, intriguingly, the CDF value of $m_W$ does lend itself to a number of new physics explanations. In this paper, we performed several different fits to electroweak observable data. 
The best description of the data is obtained with a global flavorful fit, where preference for flavor non-universality is indicated by the data. 
A subset of operators can also give a reasonable description of the data, most notably a fit that assumes the new physics resides mainly in the oblique parameters. 
The new physics contributions that affect the extraction of the electroweak VEV, such as the change in muon lifetime, on the other hand, while improving the quality of the fit compared to the SM,  cannot explain the entire tension between the CDF $m_W$ measurement and the SM prediction. 
Finally, we point out that top partners with order TeV masses could lead to the observed shift in the $W$ mass.

\subsection*{Acknowledgments}
We thank Yael Shadmi for useful discussions. The work of RB and YS is supported by grants from the NSF-BSF (No. 2018683), the ISF (No. 482/20), the BSF (No. 2020300) and by the Azrieli foundation. The work of EM is supported by the Minerva Foundation. TM and JZ acknowledge support in part by the DOE grant de-sc0011784.
The work of GP is supported by grants from BSF-NSF (No. 2019760),
Friedrich Wilhelm Bessel research award, GIF, the ISF
(grant No. 718/18), Minerva, SABRA-Yeda-Sela-WRC
Program, the Estate of Emile Mimran, and The Maurice
and Vivienne Wohl Endowment.

\appendix

\section{Fit with $m_W^{\rm PDG}$}
\label{sec:oldFit}

In this appendix we collect the numerical results of the full flavorful fit, which in contrast to the results in the main text uses the PDG value of the $W$ mass, $m_W^{\rm PDG}$ (and not $m_W^{\rm CDF}$), while all the remaining inputs are the same. This gives for the modified Wilson coefficients
\begin{gather}
    [c_{\ell\ell}]_{1221} = ( -1.07\pm 0.91 )\times 10^{-2}\,, \notag\\
    [\hat{c}'_{H\ell}]_{ii} = \begin{pmatrix}
         -0.42\pm 0.33 \\
         -0.25\pm 0.27 \\
         0.14\pm 0.41 
    \end{pmatrix} \times 10^{-2}\,,\qquad 
    [\hat{c}_{H\ell}]_{ii} = \begin{pmatrix}
         0.33\pm 0.31 \\
         0.1\pm 0.34 \\
         -0.25\pm 0.42 
    \end{pmatrix} \times 10^{-2}\,, \notag\\
    [\hat{c}_{He}]_{ii} = \begin{pmatrix}
         0.03\pm 0.06 \\
         -0.04\pm 0.27 \\
         -0.18\pm 0.13 
    \end{pmatrix} \times 10^{-2}\,, \\
    [\hat{c}'_{Hq}]_{ii} = \begin{pmatrix} 
         0.61\pm 2.67 \\
         -1.38\pm 2.78 \\
         -0.29\pm 4.10 
    \end{pmatrix} \times 10^{-2}\,,\qquad 
    [\hat{c}_{Hq}]_{ii} = \begin{pmatrix}
         3.06\pm 5.49 \\
         -0.94\pm 2.83 \\
         -0.44\pm 4.11  
    \end{pmatrix} \times 10^{-2}\,, \notag\\
    [\hat{c}_{Hu}]_{ii} = \begin{pmatrix}
         0.97\pm 6.27 \\
         0.76\pm 1.04 \\
        \text{---}
    \end{pmatrix} \times 10^{-2}\,,\qquad 
    [\hat{c}_{Hd}]_{ii} = \begin{pmatrix}
         5.11\pm 25.84 \\
         -6.42\pm 9.68 \\
         -4.5\pm 1.74 
    \end{pmatrix} \times 10^{-2}\,.\notag
\end{gather}
The fit yields $\chi^2/\text{d.o.f.} = 26.6/(42-21)$ with $\Delta\chi^2 = 17$ compared to the pure SM without dimension six contributions.
The corresponding prediction for the $W$ mass is $m_W = \SI{80.379(12)}{\GeV}$.

\section{Inputs and correlation matrices}
\label{app:inputs}

The inputs used in the global fit are given in \cref{tab:obs}. We also list the two correlation matrices, for the fit using the CDF value of $m_W$, and for the fit using the PDG value. 
The correlation matrix for the CDF fit can be found in \cref{eq:rhoCDF}. The correlation matrix following from the fit using the PDG average for the $W$ mass is given in \cref{eq:rhoPDG}.
The rows and columns correspond to the 21 effective Wilson coefficients ($[c_{\ell\ell}]_{1221}$, $[\hat{c}_{H\ell}']_{11}$, $[\hat{c}_{H\ell}']_{22}$, $[\hat{c}_{H\ell}']_{33}$, $[\hat{c}_{H\ell}]_{11}$, $[\hat{c}_{H\ell}]_{22}$, $[\hat{c}_{H\ell}]_{33}$, $[\hat{c}_{He}]_{11}$, $[\hat{c}_{He}]_{22}$, $[\hat{c}_{He}]_{33}$, $[\hat{c}_{Hq}']_{11}$, $[\hat{c}_{Hq}']_{22}$, $[\hat{c}_{Hq}']_{33}$, $[\hat{c}_{Hq}]_{11}$, $[\hat{c}_{Hq}]_{22}$, $[\hat{c}_{Hq}]_{33}$, $[\hat{c}_{Hu}]_{11}$, $[\hat{c}_{Hu}]_{22}$, $[\hat{c}_{Hd}]_{11}$, $[\hat{c}_{Hd}]_{22}$, $[\hat{c}_{Hd}]_{33}$), where for simplicity we neglected the flavor off-diagonal components in the fit.

\begin{landscape}
\begin{table}
    \centering
    \begin{tabular}{c|c|c|c||c|c|c|c}
        Observable & Experimental value & SM value & Ref. & Observable & Experimental value & SM value & Ref. \\\hline
        $m_Z$ & \SI{91.1876 \pm 0.0021}{\GeV} & \SI{91.1882}{\GeV} & \cite{ParticleDataGroup:2020ssz} &
        $m_W$ & \SI{80.4335\pm0.0094}{\GeV} & \SI{80.361}{\GeV} & \cite{CDF:mW} \\
        $\Gamma_Z$ & \SI{2.4955\pm 0.0023}{\GeV} & \SI{2.4941}{\GeV} & \cite{ALEPH:2005ab,Janot:2019oyi} &
        $\Gamma_W$ & \SI{2.085\pm0.042}{\GeV} & \SI{2.088}{\GeV} & \cite{ParticleDataGroup:2020ssz} \\
        $\sigma_\text{had}$ & \SI{41.4802\pm0.0325}{\nano\barn} & \SI{41.4842}{\nano\barn}
         & \cite{ALEPH:2005ab,Janot:2019oyi} &
        $\text{Br}\left(W\to e\nu\right)$ & \num{0.1071\pm0.0016} & \num{0.1082} & \cite{ALEPH:2013dgf} \\ 
        $R_e$ & \num{20.804\pm0.050} & \num{20.734} & \cite{ALEPH:2005ab} &
        $\text{Br}\left(W\to \mu\nu\right)$ & \num{0.1063\pm0.0015} & \num{0.1082} & \cite{ALEPH:2013dgf} \\ 
        $R_\mu$ & \num{20.785\pm0.033} & \num{20.734} & \cite{ALEPH:2005ab} &
        $\text{Br}\left(W\to \tau\nu\right)$ & \num{0.1138\pm0.0021} & \num{0.1081} & \cite{ALEPH:2013dgf} \\ 
        $R_\tau$ & \num{20.764\pm0.045} & \num{20.781} & \cite{ALEPH:2005ab} &
        $\frac{\text{Br}\left(W\to \mu\nu\right)}{\text{Br}\left(W\to e\nu\right)}$ & \num{0.982\pm0.024} & \num{1} & \cite{CDF:2005bdv} \\
        $A_\text{FB}^{0,e}$ & \num{0.0145\pm0.0025} & \num{0.0162} & \cite{ALEPH:2005ab} &
         & \num{1.020\pm0.019} & & \cite{LHCb:2016zpq} \\
        $A_\text{FB}^{0,\mu}$ & \num{0.0169\pm0.0013} & \num{0.0162} & \cite{ALEPH:2005ab} &
         & \num{1.003\pm0.010} & & \cite{ATLAS:2016nqi} \\
        $A_\text{FB}^{0,\tau}$ & \num{0.0188\pm0.0017} & \num{0.0162} & \cite{ALEPH:2005ab} &
        $\frac{\text{Br}\left(W\to \tau\nu\right)}{\text{Br}\left(W\to e\nu\right)}$ & \num{0.961\pm0.061} & \num{0.999} & \cite{ParticleDataGroup:2020ssz,D0:1999bqi} \\
        $R_b$ & \num{0.21629\pm0.00066} & \num{0.21581} & \cite{ALEPH:2005ab} &
        $\frac{\text{Br}\left(W\to \tau\nu\right)}{\text{Br}\left(W\to \mu\nu\right)}$ & \num{0.992\pm0.013}& \num{0.999} & \cite{ATLAS:2020xea} \\
        $R_c$ & \num{0.1721\pm0.0030} & \num{0.17222} & \cite{ALEPH:2005ab} &
        $R_{Wc}$ & \num{0.49\pm0.04} & \num{0.50} & \cite{ParticleDataGroup:2020ssz} \\
        $A_\text{FB}^{0,b}$ & \num{0.0996\pm0.0016} & \num{0.1032} & \cite{ALEPH:2005ab,dEnterria:2020cgt} &
        $R_\sigma$ & \num{0.998\pm0.041} & \num{1} & \cite{CMS:2014mgj} \\
        $A_\text{FB}^{0,c}$ & \num{0.0707\pm0.0035} & \num{0.0736} & \cite{ALEPH:2005ab} &
        $g_{u,A}^\text{D0}$ & \num{0.501\pm0.110} & \num{0.501} & \cite{D0:2011baz} \\
        $A_e$ & \num{0.1516\pm0.0021} & \num{0.1470} & \cite{ALEPH:2005ab} &
        $g_{u,V}^\text{D0}$ & \num{0.201\pm0.112} & \num{0.192} & \cite{D0:2011baz} \\
              & \num{0.1498\pm0.0049} & & \cite{ALEPH:2005ab} &
        $g_{d,A}^\text{D0}$ & \num{-0.497\pm0.165} & \num{-0.502} & \cite{D0:2011baz} \\
        $A_\mu$ & \num{0.142\pm0.015} & \num{0.1470} & \cite{ALEPH:2005ab} &
        $g_{d,V}^\text{D0}$ & \num{-0.351\pm0.251} & \num{-0.347} & \cite{D0:2011baz} \\
        $A_\tau$ & \num{0.136\pm0.015} & \num{0.1470} & \cite{ALEPH:2005ab} &
        $A_4(0\leq|Y|<0.8)$ & \num{0.0195\pm0.0015} & \num{0.0144} & \cite{ATLAS:2018gqq,Breso-Pla:2021qoe} \\
                 & \num{0.1439\pm0.0043} & & \cite{ALEPH:2005ab} &
        $A_4(0.8\leq|Y|<1.6)$ & \num{0.0448\pm0.0016} & \num{0.0471} & \cite{ATLAS:2018gqq,Breso-Pla:2021qoe} \\         
        $A_b$ & \num{0.923\pm0.020} & \num{0.935} & \cite{ALEPH:2005ab} &
        $A_4(1.6\leq|Y|<2.5)$ & \num{0.0923\pm0.0026} & \num{0.0928} & \cite{ATLAS:2018gqq,Breso-Pla:2021qoe} \\
        $A_c$ & \num{0.670\pm0.027} & \num{0.668} & \cite{ALEPH:2005ab} &
        $A_4(2.5\leq|Y|<3.2)$ & \num{0.1445\pm0.0046} & \num{0.1464} & \cite{ATLAS:2018gqq,Breso-Pla:2021qoe} \\
        $A_s$ & \num{0.895\pm0.091} & \num{0.936} & \cite{SLD:2000jop} &
        $R_{uc}$ & \num{0.166\pm0.009} & \num{0.1722} & \cite{ParticleDataGroup:2020ssz} \\
    \end{tabular}
    \caption{Observables used in the fit.}
    \label{tab:obs}
\end{table}

\begin{equation}
    \rho^\text{CDF} = \left(\begin{array}{ccccccccccccccccccccc}
         100 & 82 & 71 & -64 & -78 & -47 & 62 & -19 & 19 & 13 & 1 & 3 & 0 & 5 & 3 & -1 & 7 & -1 & 8 & 5 & -1 \\
         82 & 100 & 21 & -57 & -97 & -8 & 56 & -10 & 14 & 10 & 1 & 3 & 0 & 4 & 3 & 0 & 5 & -1 & 6 & 4 & -1 \\
         71 & 21 & 100 & -37 & -14 & -69 & 37 & -12 & 17 & 13 & 1 & 3 & 0 & 4 & 3 & 0 & 6 & -1 & 7 & 4 & -1 \\
         -64 & -57 & -37 & 100 & 64 & 36 & -95 & -8 & 11 & 8 & 0 & 0 & 0 & 0 & 0 & 0 & 0 & 0 & 0 & 0 & -1 \\
         -78 & -97 & -14 & 64 & 100 & 4 & -61 & 19 & -14 & -7 & 0 & -2 & 0 & -3 & -2 & 0 & -5 & 2 & -5 & -3 & -5 \\
         -47 & -8 & -69 & 36 & 4 & 100 & -32 & 13 & 53 & -10 & 0 & -1 & 0 & -3 & -2 & 1 & -4 & 0 & -5 & -3 & 3 \\
         62 & 56 & 37 & -95 & -61 & -32 & 100 & 13 & -11 & 5 & 0 & 1 & 0 & 0 & 0 & 0 & 0 & 0 & 0 & 0 & 1 \\
         -19 & -10 & -12 & -8 & 19 & 13 & 13 & 100 & -3 & 4 & 0 & -1 & 0 & -2 & 0 & -3 & -2 & 10 & -3 & -2 & -35 \\
         19 & 14 & 17 & 11 & -14 & 53 & -11 & -3 & 100 & 5 & 0 & 1 & 0 & 2 & 1 & 0 & 3 & -2 & 3 & 2 & 3 \\
         13 & 10 & 13 & 8 & -7 & -10 & 5 & 4 & 5 & 100 & 0 & 1 & 0 & 2 & 1 & 0 & 2 & -1 & 3 & 2 & 0 \\
         1 & 1 & 1 & 0 & 0 & 0 & 0 & 0 & 0 & 0 & 100 & -94 & 0 & 53 & -92 & 0 & 32 & -1 & 14 & 14 & 0 \\
         3 & 3 & 3 & 0 & -2 & -1 & 1 & -1 & 1 & 1 & -94 & 100 & 0 & -39 & 98 & 0 & -12 & -1 & 7 & -1 & 0 \\
         0 & 0 & 0 & 0 & 0 & 0 & 0 & 0 & 0 & 0 & 0 & 0 & 100 & 0 & 0 & -100 & 0 & 0 & 0 & 0 & 0 \\
         5 & 4 & 4 & 0 & -3 & -3 & 0 & -2 & 2 & 2 & 53 & -39 & 0 & 100 & -39 & 0 & 58 & 7 & 65 & -18 & -1 \\
         3 & 3 & 3 & 0 & -2 & -2 & 0 & 0 & 1 & 1 & -92 & 98 & 0 & -39 & 100 & 0 & -10 & 8 & 8 & 0 & -3 \\
         -1 & 0 & 0 & 0 & 0 & 1 & 0 & -3 & 0 & 0 & 0 & 0 & -100 & 0 & 0 & 100 & 0 & -1 & 0 & 0 & 7 \\
         7 & 5 & 6 & 0 & -5 & -4 & 0 & -2 & 3 & 2 & 32 & -12 & 0 & 58 & -10 & 0 & 100 & 7 & 95 & -39 & -2 \\
         -1 & -1 & -1 & 0 & 2 & 0 & 0 & 10 & -2 & -1 & -1 & -1 & 0 & 7 & 8 & -1 & 7 & 100 & 9 & 5 & -16 \\
         8 & 6 & 7 & 0 & -5 & -5 & 0 & -3 & 3 & 3 & 14 & 7 & 0 & 65 & 8 & 0 & 95 & 9 & 100 & -46 & -2 \\
         5 & 4 & 4 & 0 & -3 & -3 & 0 & -2 & 2 & 2 & 14 & -1 & 0 & -18 & 0 & 0 & -39 & 5 & -46 & 100 & -1 \\
         -1 & -1 & -1 & -1 & -5 & 3 & 1 & -35 & 3 & 0 & 0 & 0 & 0 & -1 & -3 & 7 & -2 & -16 & -2 & -1 & 100 
    \end{array}\right) \times 10^{-2} \label{eq:rhoCDF}
\end{equation}

\begin{equation}
    \rho^\text{PDG} = \left( \begin{array}{ccccccccccccccccccccc}
         100 & 82 & 71 & -63 & -78 & -46 & 62 & -17 & 19 & 13 & 1 & 3 & 0 & 5 & 3 & -1 & 7 & -1 & 8 & 5 & -1 \\
         82 & 100 & 21 & -56 & -97 & -8 & 56 & -7 & 14 & 11 & 1 & 3 & 0 & 4 & 3 & 0 & 5 & -1 & 6 & 4 & -1 \\
         71 & 21 & 100 & -36 & -14 & -68 & 37 & -8 & 18 & 14 & 1 & 3 & 0 & 4 & 3 & 0 & 6 & -1 & 7 & 4 & -1 \\
         -63 & -56 & -36 & 100 & 64 & 36 & -95 & -6 & 12 & 9 & 0 & 0 & 0 & 0 & 0 & 0 & 0 & 0 & 0 & 0 & -1 \\
         -78 & -97 & -14 & 64 & 100 & 4 & -61 & 19 & -13 & -6 & 0 & -2 & 0 & -3 & -2 & 0 & -5 & 2 & -5 & -3 & -5 \\
         -46 & -8 & -68 & 36 & 4 & 100 & -32 & 13 & 53 & -9 & 0 & -1 & 0 & -3 & -2 & 1 & -4 & 0 & -5 & -3 & 4 \\
         62 & 56 & 37 & -95 & -61 & -32 & 100 & 13 & -11 & 6 & 0 & 1 & 0 & 0 & 0 & 0 & 0 & 0 & 0 & 0 & 1 \\
         -17 & -7 & -8 & -6 & 19 & 13 & 13 & 100 & 0 & 8 & 1 & -1 & 1 & -2 & 0 & -3 & -2 & 9 & -2 & -2 & -33 \\
         19 & 14 & 18 & 12 & -13 & 53 & -11 & 0 & 100 & 6 & 0 & 1 & 0 & 2 & 1 & 0 & 3 & -2 & 3 & 2 & 3 \\
         13 & 11 & 14 & 9 & -6 & -9 & 6 & 8 & 6 & 100 & 0 & 1 & 0 & 2 & 1 & 0 & 2 & -1 & 3 & 2 & 0 \\
         1 & 1 & 1 & 0 & 0 & 0 & 0 & 1 & 0 & 0 & 100 & -94 & 0 & 53 & -92 & 0 & 32 & -1 & 14 & 14 & 0 \\
         3 & 3 & 3 & 0 & -2 & -1 & 1 & -1 & 1 & 1 & -94 & 100 & 0 & -39 & 98 & 0 & -12 & -1 & 7 & -1 & 0 \\
         0 & 0 & 0 & 0 & 0 & 0 & 0 & 1 & 0 & 0 & 0 & 0 & 100 & 0 & 0 & -100 & 0 & 0 & 0 & 0 & 0 \\
         5 & 4 & 4 & 0 & -3 & -3 & 0 & -2 & 2 & 2 & 53 & -39 & 0 & 100 & -39 & 0 & 58 & 7 & 65 & -18 & -1 \\
         3 & 3 & 3 & 0 & -2 & -2 & 0 & 0 & 1 & 1 & -92 & 98 & 0 & -39 & 100 & 0 & -10 & 8 & 8 & 0 & -3 \\
         -1 & 0 & 0 & 0 & 0 & 1 & 0 & -3 & 0 & 0 & 0 & 0 & -100 & 0 & 0 & 100 & 0 & -1 & 0 & 0 & 7 \\
         7 & 5 & 6 & 0 & -5 & -4 & 0 & -2 & 3 & 2 & 32 & -12 & 0 & 58 & -10 & 0 & 100 & 7 & 95 & -39 & -2 \\
         -1 & -1 & -1 & 0 & 2 & 0 & 0 & 9 & -2 & -1 & -1 & -1 & 0 & 7 & 8 & -1 & 7 & 100 & 9 & 5 & -16 \\
         8 & 6 & 7 & 0 & -5 & -5 & 0 & -2 & 3 & 3 & 14 & 7 & 0 & 65 & 8 & 0 & 95 & 9 & 100 & -46 & -2 \\
         5 & 4 & 4 & 0 & -3 & -3 & 0 & -2 & 2 & 2 & 14 & -1 & 0 & -18 & 0 & 0 & -39 & 5 & -46 & 100 & -1 \\
         -1 & -1 & -1 & -1 & -5 & 4 & 1 & -33 & 3 & 0 & 0 & 0 & 0 & -1 & -3 & 7 & -2 & -16 & -2 & -1 & 100
    \end{array} \right) \times 10^{-2} \label{eq:rhoPDG}
\end{equation}
\end{landscape}

\bibliography{mWbiblio}

\end{document}